\documentclass[%
 twocolumn,
 amsmath,amssymb,
 aps,pre,
]{revtex4-1}

\usepackage{graphicx}% Include figure files
\usepackage{dcolumn}% Align table columns on decimal point
\usepackage{bm}% bold math
\usepackage{natbib}
\begin{document}

\title{Phonon contribution to the entropy of hard sphere crystals}% Force line breaks with \\

\author{Veit Elser}
\email{ve10@cornell.edu}
\affiliation{Department of Physics\\
Cornell University
}


\begin{abstract}
Comparing the entropies of hard spheres in the limit of close packing, for different
stacking sequences of the hexagonal layers, has been a challenge because the differences
are so small. Here we present a new method based on a ``sticky-sphere" model by which the system interpolates
between hard spheres in one limit and a harmonic crystal in the other.
For the fcc and hcp stackings we have calculated the entropy difference in the harmonic (sticky) limit, as well as the differences in the free energy change upon removing the stickiness in the model. The former, or phonon entropy, accounts for most of the entropy difference. Our value for the net entropy difference, $\Delta s=0.001164(8) k_\mathrm{B}$ per sphere, is in excellent agreement with the best previous estimate by Mau and Huse [Phys. Rev. E \textbf{59}, 4396 (1999)].  
\end{abstract}

\pacs{Valid PACS appear here}% PACS, the Physics and Astronomy
\maketitle

%\tableofcontents

\section{Introduction}

The structural degeneracy of high density phases of hard spheres, owing to the stacking sequence freedom
of the hexagonal layers, is in principle resolved by differences in free-volume motion that persists even in the
limit of close-packing. As pointed out by Stillinger \textit{et al.} \cite{stillinger1} nearly 50 years ago, the free-volume and associated
entropy difference is likely
very small, since (in the close packing limit) it is exactly zero in the approximation where each sphere is constrained by neighbors that are fixed at their average positions. Resolving just the sign of the entropy difference between the
extreme cases of the fcc and hcp stackings required innovative Monte Carlo methods and substantial computing resources. Whereas there is now also good quantitative agreement in the magnitude of the entropy difference, the methods used up to now have not been able to identify, in qualitative terms, the main source of the entropy difference.

Three computational methods have been applied to the entropy difference problem. \textit{Integration methods} obtain
free energy differences by integrating thermodynamic quantities along paths that take the system from one or
the other close packed structure to a convenient reference system as a parameter is varied. Two examples are (1) differences in the integrated  pressure when the two crystal phases are expanded into the liquid phase \cite{frenkelsmit}, and (2) differences in the integrated ``Einstein oscillator" energy when fictitious Einstein couplings of the spheres to fcc or hcp sites are gradually reduced to zero from a large value \cite{frenkelladd}. In (1) the reference system (liquid) is not analytically tractable but common to the two paths, while in (2) the reference systems are distinct but have the same (analytic) free energy. The best results have been obtained by the Einstein oscillator method, the most accurate reported entropy difference being $\Delta s=0.00094(30) k_\mathrm{B}$\footnote{Figures in parentheses are the estimated uncertainty in the final digits of the quoted results.} per sphere by Bolhuis \textit{et al.} \cite{bolhuis}.

The second computational method extracts the entropy difference from the relative rate the system visits the two crystal structures, where equilibration in the presence of the large free energy barrier separating them is enabled by the \textit{multicanonical Monte Carlo} (MCMC) technique \cite{bergneuhaus1,bergneuhaus2}. In MCMC the probability distribution for the barrier configurations is modified so as to eliminate the barrier, thereby allowing the system to easily diffuse between fcc and hcp stackings. Two schemes have been tried. In the ``lattice switching" method \cite{bruce} weights are introduced that encourages the system, when it has no sphere overlaps for one reference crystal, to also have few overlaps when the reference sites are switched to the other crystal (while keeping displacements from the sites fixed). A very different method \cite{mauhuse} for eliminating the barrier is to modify the minimum pairwise distances between spheres in the space of configurations generated by applying a shear transformation to the system that interpolates between the two crystal forms. The lattice switching scheme currently provides the best estimate of the entropy difference: $\Delta s=0.00115(4) k_\mathrm{B}$. Using this method Mau and Huse \cite{mauhuse} were also able to determine the parameters in a free energy expression for general stacking sequences and confirmed that fcc and hcp are extremes in the entropy.

The third approach develops the entropy in a series not unlike the virial expansion for a liquid \cite{stillinger1,stillinger2}. Called \textit{cell cluster analysis}, first an ``unperturbed" Hamiltonian is defined having a sum of terms that exactly localizes each sphere at a crystal reference site. To recover the original hard sphere model, the negative of the localizing terms is treated as a perturbation and the free energy is expanded in a cluster series where the amplitude of the perturbation is treated as a small parameter. In the close packing limit, where constraints on the sphere positions reduce to linear inequalities, the individual cluster integrals are polytope volumes associated with the free motion of contiguous sets of spheres surrounded by fixed boundaries. The convergence properties of this series is not known, but low order results, especially in the case of disks in two dimensions, are encouraging \cite{stillinger1}. Unfortunately, even up to third order this method gives a higher entropy to the hcp structure. When the calculations were extended to two more orders \cite{koch} it was found fcc is first favored at fifth order, with an entropy difference $\Delta s=0.00115 k_\mathrm{B}$, essentially equal to what we know to be correct from the MCMC work. However, based on the sizes of the contributions to the series estimate, the uncertainty is believed to be significantly larger than the MCMC uncertainty.

The present work is based on an integration method, but differs from previous work in that the analytically tractable reference systems for the fcc and hcp structures, at infinite parameter value, already have an entropy difference $\Delta s(\infty)=0.001475 k_\mathrm{B}$ of the correct sign and about the right magnitude. Although extensive Monte Carlo sampling is still required to accurately determine the change upon reducing the parameter to zero, $\Delta s(0)-\Delta s(\infty)=-0.000311(8) k_\mathrm{B}$, the reference systems appear to have captured the relevant characteristics responsible for the entropy difference.

\section{Sticky-sphere model}\label{sec2}

Consider a hard sphere solid with average sphere centers $R_i$ at sites of a crystal with unit near-neighbor distance, and sphere positions displaced by $r_i$ relative to these sites. A pair of adjacent spheres of diameter $1-\delta$ centered at crystal sites can move a distance $\delta$ toward each other before they encounter the hard wall constraint. Instead of keeping the energy constant at zero at all separations that do not violate the constraint, we choose a quadratic energy that favors exactly the gaps of size $\delta$ between spheres. We will be working at the close-packed limit where the interaction between adjacent spheres is just a function of the projection of the sphere displacements $r_i$ onto the axes defined by adjacent crystal sites. The system Hamiltonian therefore takes the form
\begin{equation}
H_\epsilon=\sum_{(i j)}U_\epsilon\left((r_i-r_j)\cdot (R_i-R_j)\right)
\end{equation}
where the sum runs over adjacent crystal sites and $U_\epsilon$ is the ``sticky-sphere" potential of a single variable and energy parameter $\epsilon$,
\begin{equation}\label{potential}
U_\epsilon(x)=\left\{
\begin{array}{ll}
\infty,& x<-\delta\\
\epsilon((x/\delta)^2-1),& |x|\le \delta\\
0,&x>\delta.
\end{array}
\right.
\end{equation}
The attractive part of the potential has a short range and the hard sphere model is recovered when $\epsilon=0$.

The free energy $f(\epsilon)$ of a sticky-sphere system of $N$ spheres is defined by
\begin{equation}\label{freeenergy}
\beta f(\epsilon)=-\frac{1}{N}\log{Z_\epsilon},
\end{equation}
where $\beta$ is the inverse temperature and $Z_\epsilon$ is the classical partition function associated with $H_\epsilon$. From the derivative
\begin{equation}
\frac{\partial f}{\partial \epsilon}=\frac{1}{N}\left< \frac{\partial H_\epsilon}{\partial \epsilon}\right>_\epsilon=\frac{1}{N}\langle H_{\epsilon=1}\rangle_\epsilon,
\end{equation}
we can relate free energy changes to expectation values of $H_1$ in the Gibbs ensemble defined by $H_\epsilon$.
By design (see Section \ref{sec4}) the difference in the free energy derivative between the two crystals
\begin{equation}\label{deltae}
\Delta e=\frac{\partial f_\mathrm{fcc}}{\partial \epsilon}-\frac{\partial f_\mathrm{hcp}}{\partial \epsilon},
\end{equation}
vanishes exponentially when $\epsilon$ exceeds the thermal energy $\beta^{-1}$. The net change, in the fcc/hcp free energy difference, between the harmonic and hard sphere limits, is given by the integral
\begin{equation}\label{freeenergychange}
\Delta f(\infty)-\Delta f(0)=\int_0^\infty \Delta e\; d\epsilon
\end{equation}
and can be estimated by Monte Carlo sampling $\Delta e$ over a finite range of $\epsilon$. 

At infinite $\epsilon$, in the harmonic limit, the free energy difference defines a temperature-independent ``phonon entropy" difference:
\begin{equation}\label{phononentropy}
\Delta s(\infty)=-\beta\Delta f(\infty).
\end{equation}
Since the free energy at $\epsilon=0$ (the hard sphere model) is purely entropic we can express the entropy difference
\begin{eqnarray}
\Delta s(0)&=&-\beta \Delta f(0)\\
&=&-\beta \Delta f(\infty)+\left(\beta \Delta f(\infty)-\beta \Delta f(0)\right),
\end{eqnarray}
as the sum of the phonon entropy difference (\ref{phononentropy}) and the integral (\ref{freeenergychange}), which we call the ``anharmonic contribution". In the following Sections we write the anharmonic contribution as $\Delta s(0)-\Delta s(\infty)$.

\section{Phonon entropy contribution}\label{sec3}

At low temperatures the sticky sphere Hamiltonian reduces to
\begin{equation}\label{quad}
H_\epsilon\approx -6 N\epsilon+(\epsilon/\delta^2)Q(r_1,\ldots,r_N)
\end{equation}
where $Q$ is a dimensionless quadratic potential that of course depends on details of the crystal structure. To evaluate the partition function
\begin{equation}
Z_\epsilon=e^{-6N\beta \epsilon}\int_{V^N}d^3r_1\cdots d^3r_N e^{-(\beta\epsilon/\delta^2)Q(r_1,\ldots,r_N)}
\end{equation}
we first make an orthogonal change of variables to phonon amplitudes $\xi_\alpha$ such that
\begin{equation}
Q=\sum_{\alpha=1}^{3N}\kappa_\alpha \xi_\alpha^2.
\end{equation}
Three of the stiffnesses $\kappa_\alpha$, associated with translation modes
\begin{equation}
[\xi_1\;\xi_2\;\xi_3]=\frac{1}{\sqrt{N}}\sum_{i=1}^N r_i
\end{equation}
are exactly zero. The only dependence on the system volume $V$ comes from the integrals over these modes, since all other mode amplitudes are effectively cutoff in the $\epsilon\to \infty$ limit. The resulting partition function evaluates to
\begin{equation}\label{Z1}
Z_\epsilon=e^{-6N\beta \epsilon} N^{3/2}V\prod_{\kappa_\alpha\ne 0}\sqrt{\pi\frac{\delta^2}{\beta\epsilon}\cdot\frac{1}{\kappa_\alpha}},
\end{equation}
and the free energy (\ref{freeenergy}) in the $\epsilon\to\infty$ limit takes the form
\begin{equation}\label{phononfreeenergy}
-\beta f(\epsilon)\sim s_0(N,V,\beta\epsilon,\delta)-\frac{1}{2 N}\sum_{\kappa_\alpha\ne 0}\log{\kappa_\alpha},
\end{equation}
where the first term is independent of the crystal structure (stacking sequence). The phonon entropy difference (\ref{phononentropy}) is therefore
\begin{equation}
\Delta s(\infty)=-\frac{1}{2N}\left(\sum_{\mathrm{fcc}\;\kappa_\alpha}\log{\kappa_\alpha}-\sum_{\mathrm{hcp}\;\kappa_\alpha}\log{\kappa_\alpha}\right),
\end{equation}
where the sums include all the non-zero stiffnesses. Details on the calculations of the stiffnesses are provided in the Appendix. Because the harmonic free energy (\ref{phononfreeenergy}) is $\epsilon$-independent in the structure dependent part, all anharmonic contributions to the free energy difference are contained in the integral contribution (\ref{freeenergychange}).

\begin{table}[b]%The best place to locate the table environment is directly after its first reference in text
\caption{\label{tab:table1}%
Differences in the phonon and anharmonic contributions (second and third columns) to the hard sphere entropy (last column) for fcc and hcp systems of $N$ spheres. The extrapolation procedure used for the last row is described in the results section.
}
\begin{ruledtabular}
\begin{tabular}{rclc}
$N$&
$\Delta s(\infty)$&
$\Delta s(0)-\Delta s(\infty)$&
$\Delta s(0)$\\
\colrule
64 & 0.001216 & -0.004804(18) & -0.003588(18)\\
216 & 0.001378 & -0.000752(9) & 0.000626(9)\\
512 & 0.001431 & -0.000369(8) & 0.001062(8)\\
1000 & 0.001452 & -0.000319(6) & 0.001133(6)\\
$\infty$ & 0.001475 &  -0.000311(8)& 0.001164(8)\\
\end{tabular}
\end{ruledtabular}
\end{table}

\begin{figure}[t]
\includegraphics[width=3.in]{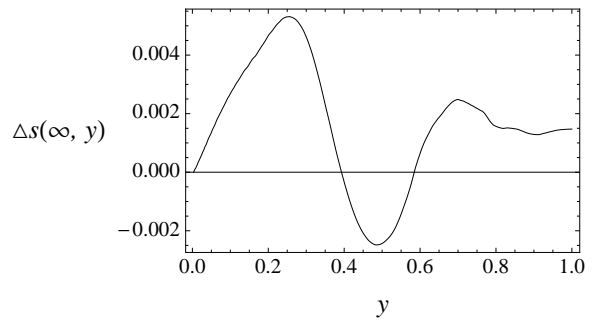}
\caption{\label{fig:fig1} Cumulative phonon entropy difference, $\Delta s(\infty,y)$, when just a fraction $y$ of the softest modes contribute. The softest and stiffest modes favor fcc, while the intermediate modes favor hcp.}
\end{figure}

\section{Anharmonic contribution}\label{sec4}

To estimate (\ref{freeenergychange}) we used the Metropolis algorithm to sample the Gibbs distribution of the sticky sphere Hamiltonian for each crystal at closely spaced values of $\epsilon$ and summed the averages $\langle H_1\rangle_\epsilon$. 
Elementary transitions were implemented by adding a displacement vector, sampled uniformly within a cube, to one of the sphere positions and accepting the transition by the Metropolis criterion. One application of the elementary transition to each of the spheres in turn comprised a ``sweep" of the system. The range of the attempted displacements was adjusted during an equilibration stage to keep the acceptance rate at approximately 50\%. At each $\epsilon$ we performed $10^8$ sweeps to find the average $\langle H_1\rangle_\epsilon$, preceded by $10^7$ sweeps to bring the system into equilibrium upon changing $\epsilon$.

Our Metropolis sampling results should be consistent with the small and large $\epsilon$ limits of $\Delta e$, the difference in the expectation value of the attractive part of the Hamiltonian in the two crystals. Since the first two $\epsilon$-derivatives of $f$ at $\epsilon=0$ are respectively the mean value and variance of $H_1$ in the hard sphere ensemble, $\Delta e$ should be linear at small $\epsilon$. At the opposite extreme, for $\epsilon\gg  k_\mathrm{B} T$, $\Delta e$ should vanish exponentially since the harmonic crystal has
\begin{equation}
\langle H_1\rangle_\epsilon/N\sim -6+(3/2)(1-1/N)k_\mathrm{B} T/\epsilon,
\end{equation}
independent of crystal structure, and is separated by a gap to the anharmonic interactions.

\begin{figure}[t]
\includegraphics[width=3.in]{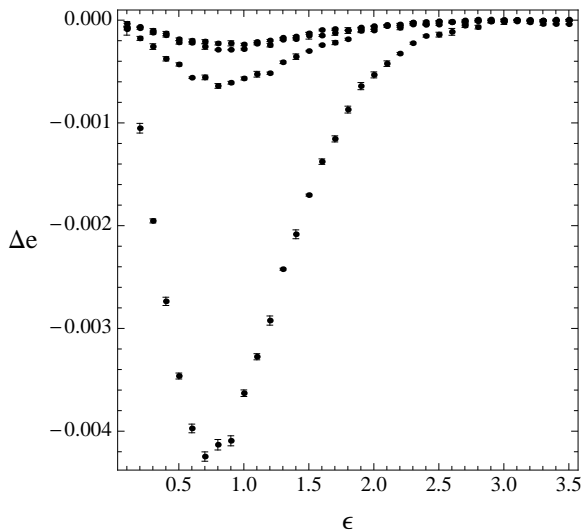}
\caption{\label{fig:fig2} Difference in the fcc and hcp free energy $\epsilon$-derivative, $\Delta e$ (\ref{deltae}), as a function of the depth of the sticky sphere potential, $\epsilon$. This quantity has a very strong system-size dependence, with $\Delta e$ of the 64-sphere systems larger by more than an order of magnitude than that of the 1000-sphere systems. Data for the two largest systems is shown in greater detail in Figure 3. The energy unit on the $\epsilon$-axis is the thermal energy $k_\mathrm{B} T$.}
\end{figure}

\begin{figure}[t]
\includegraphics[width=3.in]{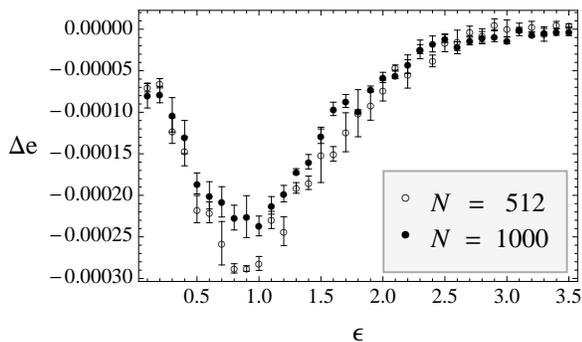}
\caption{\label{fig:fig2} Same as Figure 2, but just showing the two largest system sizes.}
\end{figure}

\section{Results}

Phonon entropy differences for the four sizes of systems simulated, along with the infinite system limit, are given in Table \ref{tab:table1}. Phonons in the fcc crystal have, on average, slightly smaller log-stiffnesses than their counterparts in the hcp crystal. However, the relative shifts of these phonon spectra is far from monotonic. This is shown in Figure \ref{fig:fig1}, where the entropy difference $\Delta s(\infty,y)$, due to just the softest fraction $y$ of the modes, is plotted vs. $y$.

The integral of $\beta\Delta e$, which is the change in the entropy difference (between hard spheres and harmonic crystal) $\Delta s(0)-\Delta s(\infty)$, is also recorded in Table \ref{tab:table1}. It is interesting that while these ``corrections" to the entropy difference have a significant size effect for the smallest system ($N=64$), the phonon entropy difference $\Delta s(\infty)$ itself does not. 

In Figures 2 and 3 we plot $\Delta e$ as a function of $\epsilon$, where the energy unit on the $\epsilon$ axis is $\beta^{-1}=k_\mathrm{B} T$. The limiting behaviors at small and large $\epsilon$, discussed in Section \ref{sec4}, are both borne out by the data.

Given the small value of the fcc/hcp entropy difference, even in the close packing limit, it is not surprising that our results are sensitive to boundary conditions. A case in point is the entropy difference for our systems of 64 spheres, where for our choice of boundary conditions the hcp structure is strongly favored ($\Delta s(0)=-0.0036 k_\mathrm{B}$). Mau and Huse \cite{mauhuse} use a lower symmetry boundary condition for their hcp systems and obtain a much weaker system size dependence, where fcc is favored for all sizes. In the 1000-sphere systems the boundary effects have become sufficiently small that our entropy difference is in perfect agreement with theirs.

We applied different extrapolation procedures to the phonon and anharmonic contributions to arrive at a thermodynamic limit estimate of the entropy difference. The integral for the phonon entropy has a logarithmic non-analyticity at zero momentum, and finite size corrections should decay with the linear scale $n=N^{1/3}$ as  $\log{n}/n^3$. This form is consistent with our calculations (up to $n=24$) and gives an extrapolated phonon entropy difference $\Delta s(\infty)=0.001475 k_\mathrm{B}$. Finite size corrections for the anharmonic contribution are not so easily analyzed. Excitations here have a more discrete character, associated with pairs of spheres that have escaped the attractive part of the potential and become unbound. The distribution in space of such unbound pairs should have a finite correlation length $\xi$, and finite size effects should decay exponentially on this scale. Fitting the correction to the anharmonic contributions for our three largest sizes ($N=6^3, 8^3, 10^3$) to the form $\exp{(-n/\xi)}$, we obtain $\xi\approx 1$ and the extrapolated value $\Delta s(0)-\Delta s(\infty)=-0.000311(8) k_\mathrm{B}$. Combining the phonon and anharmonic contributions, our estimate for the net entropy difference is $\Delta s(0)=0.001164(8) k_\mathrm{B}$ per sphere.

\section{Discussion}

How does our phonon method compare with previous methods? For deciding which crystal at close packing has the greatest entropy, the size of our statistical error would argue that this could have been done when computers were about $10^4$ times slower than the computer used in the present study. From this one might argue that this question could have been resolved even without the multicanonical Monte Carlo methods. Additionally, the phonon method is useful in that it identifies an analytically quantifiable source that accounts for most of the entropy difference. On the other hand, it would be going too far to use the phonon contribution as a proxy for the entropy of the hard sphere system. A proxy of some kind, that could be calculated efficiently, would be useful for properties such as the entropic elastic constants or the deviation of the equilibrium hcp solid from the ``ideal" hexagonal crystal parameters ($c/a=\sqrt{8/3}$). The problem with the harmonic part of the sticky-sphere model serving as a proxy is that there is no unique or ``natural" definition of the model.

As an example of the freedom one has in defining the sticky-sphere Hamiltonian, the curvature of the potential (\ref{potential}) can be modified without changing the position of the minimum. This has no effect on the phonon entropy difference when the same modification is applied to all pairs of spheres (the change will be reflected in the anharmonic contribution, before integration). But in the case of a sheared crystal (when calculating elastic constants) or the lower symmetry hexagonal crystal, there is no reason that the same curvature be given to the potential for all sphere pairs, and the phonon contribution to the entropy will be changed accordingly. Entropy differences based just on the phonon contribution are thus unreliable, unless a compelling argument could be made that singled out a particular stacking-dependent Hamiltonian.

\appendix

\section{Crystal sites and momentum samples}

The lattices $\Lambda_\mathrm{fcc}$, $\Lambda_\mathrm{hcp}$ associated with the fcc and hcp crystals are constructed as integer combinations of the rows of the following generator matrices:
\begin{eqnarray}
G_\mathrm{fcc}&=&\left[
\begin{array}{ccc}
1&0&0\\
1/2&\sqrt{3/4}&0\\
1/2&\sqrt{1/12}&\sqrt{2/3}
\end{array}
\right]\\
G_\mathrm{hcp}&=&\left[
\begin{array}{ccc}
1&0&0\\
1/2&\sqrt{3/4}&0\\
0&0&\sqrt{8/3}
\end{array}
\right].
\end{eqnarray}
The fcc crystal sites are just the lattice points $\Lambda_\mathrm{fcc}$, while for the hcp crystal we take $\Lambda_\mathrm{hcp}\cup(\Lambda_\mathrm{hcp}+v)$, where
\begin{equation}
v=\left[
\begin{array}{ccc}
1/2&\sqrt{1/12}&\sqrt{2/3}
\end{array}
\right].
\end{equation}
For both crystals we specify the periodic cell as the superlattice $\tilde{\Lambda}$ generated by
\begin{equation}
\widetilde{G}=\mathrm{diag}(a,b,c)\cdot G.
\end{equation}
The sites $R$ of the periodic fcc system are then the equivalence classes $\Lambda_\mathrm{fcc}/\tilde{\Lambda}_\mathrm{fcc}$, while for the periodic hcp system we take the union of $\Lambda_\mathrm{hcp}/\tilde{\Lambda}_\mathrm{hcp}$ and its translation by $v$. The number of sites in the periodic systems is $N=a b c$ for fcc and $N=2 a' b' c'$ for hcp. The fcc systems we studied all had $a=b=c$ ; the corresponding hcp systems (with the same number of spheres) had $a'=a$, $b'=b$, and $c'=c/2$.

The reciprocal lattice $\Lambda^\ast$ associated with $\Lambda$ is generated by the rows of the matrix $G^\ast=(G^{-1})^\mathrm{T}$. In the normal mode analysis of the harmonic Hamiltonian, mode-momenta $K$ differing by an element of $\Lambda^\ast$ are equivalent. For modes to have periodicity consistent with the superlattice $\tilde{\Lambda}$ they must be elements of the reciprocal lattice $\tilde{\Lambda}^\ast$ generated by the rows of
\begin{equation}
\widetilde{G}^\ast=(\widetilde{G}^{-1})^\mathrm{T}=\mathrm{diag}(a^{-1},b^{-1},c^{-1})\cdot G^\ast.
\end{equation}
The set of normal mode momenta $K$ is therefore identified with the equivalence classes $\tilde{\Lambda}^\ast/\Lambda^\ast$; the number of these is $a b c$ for our choice of superlattices.

In our normal mode analysis we index the sphere displacements $r$ by their associated lattice sites $R$ rather than the crystal sites. The displacements have the following expansion in terms of normal mode amplitudes $q$:
\begin{equation}\label{modeexp}
r_R=\frac{1}{\sqrt{a b c}}\sum_{K\in \tilde{\Lambda}^\ast/\Lambda^\ast} e^{i 2\pi K\cdot R} q_K.
\end{equation}
For each $K$ there are three modes, one for each vector component of $q_K$. Expression (\ref{modeexp}) serves for the sphere displacements at the sites $\Lambda_\mathrm{hcp}$ of the hcp crystal; for the spheres at the sites $\Lambda_\mathrm{hcp}+v$ we have a second set of displacements and mode amplitudes (associated with the same $R$ and $K$),
\begin{equation}\label{modeexp'}
s_R=\frac{1}{\sqrt{a b c}}\sum_{K\in \tilde{\Lambda}^\ast/\Lambda^\ast} e^{i 2\pi K\cdot R} p_K,
\end{equation}
and therefore six modes for each momentum $K$. The $3N$ scalar mode amplitudes are denoted $\xi_\alpha$ in Section \ref{sec3}.

The two crystals have different sets of vectors for the separations of adjacent spheres. For fcc there is a set of six vectors $A_0$
\begin{equation}
\begin{array}{ccccc}
\pm [&1&0&0 &]\\
\pm [&1/2&\sqrt{3/4}&0 &]\\
\pm [&1/2&-\sqrt{3/4}&0 &]
\end{array}
\end{equation}
for adjacent pairs in the same hexagonal layer and another six $A_1$
\begin{equation}
\begin{array}{ccccc}
\pm [&1/2&\sqrt{1/12}&\sqrt{2/3}&]\\
\pm [&-1/2&\sqrt{1/12}&\sqrt{2/3}&]\\
\pm [&0&-\sqrt{1/3}&\sqrt{2/3}&]
\end{array}
\end{equation}
between spheres in adjacent layers. The hcp crystal has the same set of adjacency vectors $A_0$ as the fcc crystal for spheres in the same coset of $\Lambda_\mathrm{hcp}$ and the  different set $A_1'$
\begin{equation}
\begin{array}{ccccc}
\, [&1/2&\sqrt{1/12}&\pm \sqrt{2/3}&]\\
\, [&-1/2&\sqrt{1/12}&\pm \sqrt{2/3}&]\\
\, [&0&-\sqrt{1/3}&\pm \sqrt{2/3}&]
\end{array}
\end{equation}
between different cosets (layers).

\section{fcc stiffness eigenvalues}

Substituting the mode expansion (\ref{modeexp}) into the quadratic part $Q$ of the sticky-sphere Hamiltonian we obtain
\begin{equation}
Q=\sum_{K\in \tilde{\Lambda}_\mathrm{fcc}^\ast/\Lambda_\mathrm{fcc}^\ast} q_K\cdot Q(K)\cdot q_K^\mathrm{T},
\end{equation}
where
\begin{equation}
Q(K)=\sum_{\delta R\in A_0\cup A_1}(1-\cos{K\cdot \delta R})\;\delta R^\mathrm{T} \cdot \delta R
\end{equation}
is a real-symmetric $3\times 3$ matrix and the sum is over the 12 adjacency vectors of the fcc crystal. The $3(N-1)$ nonzero eigenvalues of these matrices, for $K\ne 0$, are the eigenvalues $\kappa_\alpha$ of Section \ref{sec3}.

\section{hcp stiffness eigenvalues}

In the hcp case we use (\ref{modeexp}) and (\ref{modeexp'}) for the two kinds of layers, the adjacency vectors $A_0$ for interacting spheres in the same layer and vectors $A_1'$ for spheres in adjacent layers. The quadratic function $Q$ now takes the form
\begin{equation}
Q=\sum_{K\in \tilde{\Lambda}_\mathrm{hcp}^\ast/\Lambda_\mathrm{hcp}^\ast}
\left[
\begin{array}{cc}
q_K& p_K
\end{array}
\right]
\cdot
\left[
\begin{array}{cc}
Q_0(K) & Q_1(K)\\
Q_1^\dag(K)& Q_0(K)
\end{array}
\right]
\cdot
\left[
\begin{array}{c}
q_K^\mathrm{T}\\
p_K^\mathrm{T}
\end{array}
\right],
\end{equation}
where
\begin{eqnarray}
Q_0(K)&=&\sum_{\delta R\in A_0}(1-\cos{K\cdot \delta R})\;\delta R^\mathrm{T} \cdot \delta R\\
&+&\sum_{\delta R\in A_1'}\delta R^\mathrm{T} \cdot \delta R
\end{eqnarray}
and
\begin{equation}
Q_1(K)=\sum_{\delta R\in A_1'}e^{i 2\pi K\cdot \delta R}\;\delta R^\mathrm{T} \cdot \delta R
\end{equation}
are $3\times 3$ Hermitian matrices. The $3(N-1)$ non-zero eigenvalues $\kappa_\alpha$ now arise as $N/2$ blocks of six, with the exception of the $K=0$ block which has the three zero eigenvalues.

\nocite{*}

\bibliography{phonon_entropy}% Produces the bibliography via BibTeX.

\providecommand{\noopsort}[1]{}\providecommand{\singleletter}[1]{#1}%
\begin{thebibliography}{11}%
\makeatletter
\providecommand \@ifxundefined [1]{%
 \@ifx{#1\undefined}
}%
\providecommand \@ifnum [1]{%
 \ifnum #1\expandafter \@firstoftwo
 \else \expandafter \@secondoftwo
 \fi
}%
\providecommand \@ifx [1]{%
 \ifx #1\expandafter \@firstoftwo
 \else \expandafter \@secondoftwo
 \fi
}%
\providecommand \natexlab [1]{#1}%
\providecommand \enquote  [1]{``#1''}%
\providecommand \bibnamefont  [1]{#1}%
\providecommand \bibfnamefont [1]{#1}%
\providecommand \citenamefont [1]{#1}%
\providecommand \href@noop [0]{\@secondoftwo}%
\providecommand \href [0]{\begingroup \@sanitize@url \@href}%
\providecommand \@href[1]{\@@startlink{#1}\@@href}%
\providecommand \@@href[1]{\endgroup#1\@@endlink}%
\providecommand \@sanitize@url [0]{\catcode `\\12\catcode `\$12\catcode
  `\&12\catcode `\#12\catcode `\^12\catcode `\_12\catcode `\%12\relax}%
\providecommand \@@startlink[1]{}%
\providecommand \@@endlink[0]{}%
\providecommand \url  [0]{\begingroup\@sanitize@url \@url }%
\providecommand \@url [1]{\endgroup\@href {#1}{\urlprefix }}%
\providecommand \urlprefix  [0]{URL }%
\providecommand \Eprint [0]{\href }%
\providecommand \doibase [0]{http://dx.doi.org/}%
\providecommand \selectlanguage [0]{\@gobble}%
\providecommand \bibinfo  [0]{\@secondoftwo}%
\providecommand \bibfield  [0]{\@secondoftwo}%
\providecommand \translation [1]{[#1]}%
\providecommand \BibitemOpen [0]{}%
\providecommand \bibitemStop [0]{}%
\providecommand \bibitemNoStop [0]{.\EOS\space}%
\providecommand \EOS [0]{\spacefactor3000\relax}%
\providecommand \BibitemShut  [1]{\csname bibitem#1\endcsname}%
\let\auto@bib@innerbib\@empty
%</preamble>
\bibitem [{\citenamefont {{Stillinger Jr.}}\ \emph {et~al.}(1965)\citenamefont
  {{Stillinger Jr.}}, \citenamefont {Salsburg},\ and\ \citenamefont
  {Kornegay}}]{stillinger1}%
  \BibitemOpen
  \bibfield  {author} {\bibinfo {author} {\bibfnamefont {F.~H.}\ \bibnamefont
  {{Stillinger Jr.}}}, \bibinfo {author} {\bibfnamefont {Z.~W.}\ \bibnamefont
  {Salsburg}}, \ and\ \bibinfo {author} {\bibfnamefont {R.~L.}\ \bibnamefont
  {Kornegay}},\ }\href@noop {} {\bibfield  {journal} {\bibinfo  {journal} {J.\
  Chem.\ Phys.}\ }\textbf {\bibinfo {volume} {43}},\ \bibinfo {pages} {932}
  (\bibinfo {year} {1965})}\BibitemShut {NoStop}%
\bibitem [{\citenamefont {Frenkel}\ and\ \citenamefont
  {Smit}(1996)}]{frenkelsmit}%
  \BibitemOpen
  \bibfield  {author} {\bibinfo {author} {\bibfnamefont {D.}~\bibnamefont
  {Frenkel}}\ and\ \bibinfo {author} {\bibfnamefont {B.}~\bibnamefont {Smit}},\
  }\href@noop {} {\emph {\bibinfo {title} {Understanding Molecular Simulation:
  From Algorithms to Applications}}}\ (\bibinfo  {publisher} {Academic},\
  \bibinfo {address} {Boston},\ \bibinfo {year} {1996})\BibitemShut {NoStop}%
\bibitem [{\citenamefont {Frenkel}\ and\ \citenamefont
  {Ladd}(1984)}]{frenkelladd}%
  \BibitemOpen
  \bibfield  {author} {\bibinfo {author} {\bibfnamefont {D.}~\bibnamefont
  {Frenkel}}\ and\ \bibinfo {author} {\bibfnamefont {A.~J.~C.}\ \bibnamefont
  {Ladd}},\ }\href@noop {} {\bibfield  {journal} {\bibinfo  {journal} {J.\
  Chem.\ Phys.}\ }\textbf {\bibinfo {volume} {81}},\ \bibinfo {pages} {3188}
  (\bibinfo {year} {1984})}\BibitemShut {NoStop}%
\bibitem [{Note1()}]{Note1}%
  \BibitemOpen
  \bibinfo {note} {Figures in parentheses are the estimated uncertainty in the
  final digits of the quoted results.}\BibitemShut {Stop}%
\bibitem [{\citenamefont {Bolhuis}\ \emph {et~al.}(1997)\citenamefont
  {Bolhuis}, \citenamefont {Frenkel}, \citenamefont {Mau},\ and\ \citenamefont
  {Huse}}]{bolhuis}%
  \BibitemOpen
  \bibfield  {author} {\bibinfo {author} {\bibfnamefont {P.~G.}\ \bibnamefont
  {Bolhuis}}, \bibinfo {author} {\bibfnamefont {D.}~\bibnamefont {Frenkel}},
  \bibinfo {author} {\bibfnamefont {S.-C.}\ \bibnamefont {Mau}}, \ and\
  \bibinfo {author} {\bibfnamefont {D.~A.}\ \bibnamefont {Huse}},\ }\href@noop
  {} {\bibfield  {journal} {\bibinfo  {journal} {Nature}\ }\textbf {\bibinfo
  {volume} {388}},\ \bibinfo {pages} {235} (\bibinfo {year}
  {1997})}\BibitemShut {NoStop}%
\bibitem [{\citenamefont {Berg}\ and\ \citenamefont
  {Neuhaus}(1991)}]{bergneuhaus1}%
  \BibitemOpen
  \bibfield  {author} {\bibinfo {author} {\bibfnamefont {B.~A.}\ \bibnamefont
  {Berg}}\ and\ \bibinfo {author} {\bibfnamefont {T.}~\bibnamefont {Neuhaus}},\
  }\href@noop {} {\bibfield  {journal} {\bibinfo  {journal} {Phys.\ Lett. B}\
  }\textbf {\bibinfo {volume} {267}},\ \bibinfo {pages} {249} (\bibinfo {year}
  {1991})}\BibitemShut {NoStop}%
\bibitem [{\citenamefont {Berg}\ and\ \citenamefont
  {Neuhaus}(1992)}]{bergneuhaus2}%
  \BibitemOpen
  \bibfield  {author} {\bibinfo {author} {\bibfnamefont {B.~A.}\ \bibnamefont
  {Berg}}\ and\ \bibinfo {author} {\bibfnamefont {T.}~\bibnamefont {Neuhaus}},\
  }\href@noop {} {\bibfield  {journal} {\bibinfo  {journal} {Phys.\ Rev.
  Lett.}\ }\textbf {\bibinfo {volume} {68}},\ \bibinfo {pages} {9} (\bibinfo
  {year} {1992})}\BibitemShut {NoStop}%
\bibitem [{\citenamefont {Bruce}\ \emph {et~al.}(1997)\citenamefont {Bruce},
  \citenamefont {Wilding},\ and\ \citenamefont {Ackland}}]{bruce}%
  \BibitemOpen
  \bibfield  {author} {\bibinfo {author} {\bibfnamefont {A.~D.}\ \bibnamefont
  {Bruce}}, \bibinfo {author} {\bibfnamefont {N.~B.}\ \bibnamefont {Wilding}},
  \ and\ \bibinfo {author} {\bibfnamefont {G.~J.}\ \bibnamefont {Ackland}},\
  }\href@noop {} {\bibfield  {journal} {\bibinfo  {journal} {Phys.\ Rev.
  Lett.}\ }\textbf {\bibinfo {volume} {79}},\ \bibinfo {pages} {3002} (\bibinfo
  {year} {1997})}\BibitemShut {NoStop}%
\bibitem [{\citenamefont {Mau}\ and\ \citenamefont {Huse}(1999)}]{mauhuse}%
  \BibitemOpen
  \bibfield  {author} {\bibinfo {author} {\bibfnamefont {S.-C.}\ \bibnamefont
  {Mau}}\ and\ \bibinfo {author} {\bibfnamefont {D.~A.}\ \bibnamefont {Huse}},\
  }\href@noop {} {\bibfield  {journal} {\bibinfo  {journal} {Phys.\ Rev. E}\
  }\textbf {\bibinfo {volume} {59}},\ \bibinfo {pages} {4396} (\bibinfo {year}
  {1999})}\BibitemShut {NoStop}%
\bibitem [{\citenamefont {Rudd}\ \emph {et~al.}(1968)\citenamefont {Rudd},
  \citenamefont {Salsburg}, \citenamefont {Yu},\ and\ \citenamefont
  {{Stillinger Jr.}}}]{stillinger2}%
  \BibitemOpen
  \bibfield  {author} {\bibinfo {author} {\bibfnamefont {W.~G.}\ \bibnamefont
  {Rudd}}, \bibinfo {author} {\bibfnamefont {Z.~W.}\ \bibnamefont {Salsburg}},
  \bibinfo {author} {\bibfnamefont {A.~P.}\ \bibnamefont {Yu}}, \ and\ \bibinfo
  {author} {\bibfnamefont {F.~H.}\ \bibnamefont {{Stillinger Jr.}}},\
  }\href@noop {} {\bibfield  {journal} {\bibinfo  {journal} {J.\ Chem.\ Phys.}\
  }\textbf {\bibinfo {volume} {49}},\ \bibinfo {pages} {4857} (\bibinfo {year}
  {1968})}\BibitemShut {NoStop}%
\bibitem [{\citenamefont {Koch}\ \emph {et~al.}(2005)\citenamefont {Koch},
  \citenamefont {Radin},\ and\ \citenamefont {Sadun}}]{koch}%
  \BibitemOpen
  \bibfield  {author} {\bibinfo {author} {\bibfnamefont {H.}~\bibnamefont
  {Koch}}, \bibinfo {author} {\bibfnamefont {C.}~\bibnamefont {Radin}}, \ and\
  \bibinfo {author} {\bibfnamefont {L.}~\bibnamefont {Sadun}},\ }\href@noop {}
  {\bibfield  {journal} {\bibinfo  {journal} {Phys.\ Rev. E}\ }\textbf
  {\bibinfo {volume} {72}},\ \bibinfo {pages} {016708} (\bibinfo {year}
  {2005})}\BibitemShut {NoStop}%
\end{thebibliography}%

\end{document}